\PassOptionsToPackage{a4paper, margin=2cm}{geometry}

\documentclass[sn-nature]{sn-jnl}

\UseRawInputEncoding 
\usepackage[utf8]{inputenc}

\usepackage{graphicx}%
\usepackage{multirow}%
\usepackage{amsmath,amssymb,amsfonts}%
\usepackage{amsthm}%
\usepackage{mathrsfs}%
\usepackage[title]{appendix}%
\usepackage{xcolor}%
\usepackage{textcomp}%
\usepackage{manyfoot}%
\usepackage{booktabs}%
\usepackage{algorithm}%
\usepackage{algorithmicx}%
\usepackage{algpseudocode}%
\usepackage{listings}%

\theoremstyle{thmstyleone}%
\theoremstyle{thmstyletwo}%

\theoremstyle{thmstylethree}%

\begin{document}

\title[Collective dynamics behind success]{Collective dynamics behind success}


\author*[1]{\fnm{Manuel S.} \sur{Mariani}}\email{manuel.mariani@business.uzh.ch}
\author[2]{\fnm{Federico} \sur{Battiston}}
\author[3,4,5]{\fnm{Emőke-Ágnes} \sur{Horvát}}
\author[6,7,8]{\fnm{Giacomo} \sur{Livan}}
\author[9]{\fnm{Federico} \sur{Musciotto}}
\author[10,11,12,4]{\fnm{Dashun} \sur{Wang}}

\affil*[1]{\orgdiv{URPP Social Networks}, \orgname{University of Zurich}, \postcode{CH-8050 Zurich}, \country{Switzerland}}
\affil[2]{\orgdiv{Department of Network and Data Science}, \orgname{Central European University}, \postcode{Vienna}, \country{Austria}}
\affil[3]{\orgdiv{School of Communication}, \orgname{Northwestern University}, \postcode{Evanston, IL}, \country{USA}}
\affil[4]{\orgdiv{McCormick School of Engineering}, \orgname{Northwestern University}, \postcode{Evanston, IL}, \country{USA}}
\affil[5]{\orgdiv{Northwestern Institute on Complex Systems}, \orgname{Northwestern University}, \postcode{Evanston, IL}, \country{USA}}
\affil[6]{\orgdiv{Dipartimento di Fisica}, \orgname{Università degli Studi di Pavia}, \postcode{27100 Pavia}, \country{Italy}}
\affil[7]{\orgdiv{Istituto Nazionale di Fisica Nucleare, Sezione di Pavia}, \postcode{27100 Pavia}, \country{Italy}}
\affil[8]{\orgdiv{Department of Computer Science}, \orgname{University College London}, \postcode{London WC1E 6EA}, \country{UK}}
\affil[9]{\orgdiv{Department of Physics and Chemistry}, \orgname{University of Palermo}, \postcode{I-90128 Palermo}, \country{Italy}}
\affil[10]{\orgdiv{Center for Science of Science and Innovation}, \orgname{Northwestern University}, \postcode{Evanston, IL}, \country{USA}}
\affil[11]{\orgdiv{Ryan Institute on Complexity}, \orgname{Northwestern University}, \postcode{Evanston, IL}, \country{USA}}
\affil[12]{\orgdiv{Kellogg School of Management}, \orgname{Northwestern University}, \postcode{Evanston, IL}, \country{USA}}


\abstract{Understanding the collective dynamics behind the success of ideas, products, behaviors, and social actors is critical for decision-making across diverse contexts, including hiring, funding, career choices, and the design of interventions for social change. Methodological advances and the increasing availability of big data now allow for a broader and deeper understanding of the key facets of success. Recent studies unveil regularities beneath the collective dynamics of success, pinpoint underlying mechanisms, and even enable predictions of success across diverse domains, including science, technology, business, and the arts. However, this research also uncovers troubling biases that challenge meritocratic views of success. This review synthesizes the growing, cross-disciplinary literature on the collective dynamics behind success and calls for further research on cultural influences, the origins of inequalities, the role of algorithms in perpetuating them, and experimental methods to further probe causal mechanisms behind success.
Ultimately, these efforts may help to better align success with desired societal values.}

\maketitle



Many key aspects of success in social and economic systems stem from collective dynamics: millions of interconnected consumers determine the success of songs, books, and businesses~\cite{yucesoy2018success,barabasi2018formula}; citizens’ decisions collectively determine the prevalence of socially beneficial behaviors~\cite{centola2021change,airoldi2024induction}; online users' sharing choices drive the virality of content~\cite{vosoughi2018spread}; networks of employees and recruiters influence the success of job seekers~\cite{rajkumar2022causal}; and the interactions among team members shape a team’s success and failure in executing complex tasks~\cite{almaatouq2020adaptive,brackbill2020impact}. In all these cases, whether involving an idea, product, behavior, individual, or team, success is not an isolated event; rather, it is the result of the choices and actions by interconnected actors at different scales.

\begin{figure*}[t]
        \centering
 \includegraphics[scale=0.8]{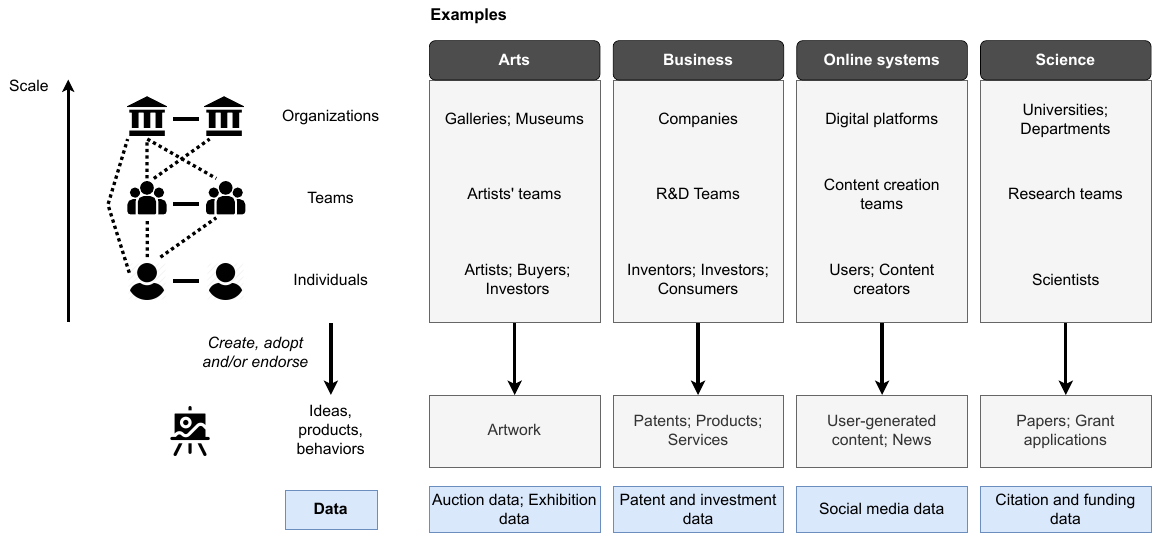}
\caption{\textbf{Success in diverse domains and at different scales}.
The recent literature on success has examined subjects across disparate domains and at different scales. 
From bottom to top, the subjects involved can be classified as individuals, teams, or large-scale organizations; these subjects create, adopt, or endorse ideas, products, and behaviors.
Recent studies on success often link subjects within scales (solid lines) and across scales (dashed lines). This review presents several instances in which these connections predict or influence individuals', teams', and organizations' success. For example, the network of interactions among potential adopters affects the successful proliferation of new ideas, products, and behaviors; the network of interactions between a team’s members affects the team’s success; and artists' and scientists' early connections with prestigious institutions predict their success.
Each gray box provides examples of individuals, teams, and organizations of interest at different scales for a given domain (from left to right: arts, business, online systems, science), as well as examples of the datasets used in the analysis (blue boxes). \textit{Icon credits:} The icons representing organizations and teams have been realized by user Freepik from \url{Flaticon.com}; The icons representing individuals and products have been realized by users Kiranshastry and bsd, respectively, from \url{Flaticon.com}.
 }
        \label{fig:framework}
    \end{figure*}

The deluge of big data capturing individual behavior and successes offers new opportunities
to explore quantitative patterns governing the collective dynamics behind success. These opportunities have stimulated the emergence of
a multidisciplinary community of scientists who are united in their quest to understand the
complex dynamics behind success across diverse socioeconomic systems, asking a wide range of fascinating questions. For example,
what makes online content go viral? 
Why is there such persistence of misinformation on social media? How
soon can we tell if a new idea or product has sparked serious interest?
When do scientists or artists do
their greatest work?
How do network structures and interventions influence the large-scale adoption of new products and behaviors, as well as the success of individual careers? 
How
is a winning team assembled? What is the role of gender in recognition and success? 
And ultimately, how can people and organizations profit from early failures to achieve long-term success?
Addressing these questions allows researchers to uncover empirical regularities that govern
the collective dynamics behind success, identify mechanisms responsible for these
patterns, and reveal predictive signals that hold practical implications. By combining diverse
approaches, results from this research program can lead--and, in several cases, have led--to qualitative
shifts in the way that new ideas are discovered, talents are identified and nurtured, excellence is
recognized and rewarded, and failure is avoided and exploited. Here, we review recent progress in this transdisciplinary research agenda.

Many of these questions have been explored in the psychology, sociology, and management literature for decades. Research in psychology, for example, has examined the lifecycles of prominent artists and scientists~\cite{simonton2004creativity,lehman2017age} and has investigated the various drivers for individual success~\cite{dweck2006mindset,duckworth2007grit} as well as team performance~\cite{hackman1975group}. Sociologists and management scholars have extensively studied the importance of various structural forces that shape career success~\cite{granovetter1973strength,burt2004structural} and the proliferation of new products and behaviors~\cite{granovetter1978threshold,bass1969new,rogers2010diffusion}.
This review builds on these foundational streams of research while highlighting recent studies that diverge from traditional approaches by combining quantitative methods with large-scale datasets drawn from real-world observations and experiments. 
This windfall of data comes from a remarkably wide range of areas, capturing patterns of success across arts, sciences, business, and more (Fig.~\ref{fig:framework}). As a result, in addition to revealing generalizable regularities and predictive signals, these efforts have validated long-standing theories at scale, delimited their boundaries to validity more accurately, and inspired the development of new theories.
Further, rapid advances in fields such as complexity science, network science, artificial intelligence, econometrics, and experimental sociology have provided powerful empirical tools and methodologies, offering deeper insights into the collective dynamics of success.

\section{The collective nature of success}

Efforts to harness the newly available data and methodologies to understand the dynamics of success encounter several core challenges.
First, the definition of success is inherently multifaceted and varies across individuals, domains, and cultures. 
The value of a work of art, for example, is subjective. Quantitative measures, such as auction prices~\cite{galenson2011old,liu2018hot,liu2021understanding}, museum exhibitions~\cite{galenson2011old,fraiberger2018quantifying}, or textbook features~\cite{galenson2011old}, are imperfect approximations for the success of the artwork. Moreover, different artists may have different career aspirations, such that one measure of artistic success may mean vastly more to a given artist.

Second, behavioral research shows that human judgment is strongly influenced by a wide range of biases and heuristics~\cite{kahneman2011thinking}, potentially impairing our ability to understand why some products, individuals, and teams succeed or fail~\cite{lifchits2021success}. 
Individuals attempting to predict or explain success may, for example, become overconfident in their predictions due to the accuracy of their previous prognostications~\cite{hilary2006does}; they may overweight extreme events while overlooking the base rate for success~\cite{denrell2010predicting}; they may fail to notice unfairness from which they benefited in their success ~\cite{molina2019s}; they may be influenced by emotionally satisfying but misleading success stories~\cite{lifchits2021success}; they may believe that success was predictable after the fact even when, a priori, it was not~\cite{fischhoff1975knew}; or they may selectively look for evidence that confirms their prior beliefs about success~\cite{nickerson1998confirmation}, among many other biases~\cite{kahneman2011thinking}.

The emerging strand of literature examines whether these challenges can be mitigated with the new availability of substantial datasets that capture performance and success at a larger scale, higher resolution, and greater level of detail than ever before, improving our understanding of success.
Here, we broadly define success as the attainment of highly-valued outcomes for an individual, team, or organization -- and for the ideas, products, or behaviors which they create or adopt.
The recent literature focuses largely on collective dimensions of success, recognizing that, in many cases, success is shaped by collective dynamics~\cite{barabasi2018formula}, both in terms of creation (the processes through which an individual, team, or organization acquires resources and generates new ideas, products, or behaviors) and in terms of reception (the processes through which a completed product or well-defined behavior is potentially adopted by interconnected individuals). 
Under this lens, understanding success requires uncovering the complex processes involving interacting agents which drive the collective adoption of new ideas, products, and behaviors, and, therefore, shape the success of individuals, teams, and organizations (see Fig.~\ref{fig:framework}).
While this focus on the collective dynamics behind success necessarily neglects some personal dimensions of success, the shift in perspective allows researchers to bypass the need to evaluate individual beliefs about success and focus instead on collective measures that are quantifiable in large-scale data sets.

\begin{figure}[t]
        \centering
 \includegraphics[scale=1.25]{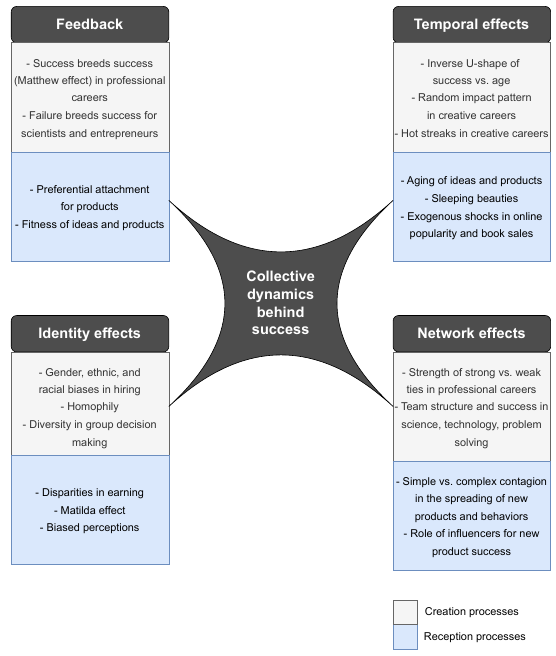}
\caption{\textbf{Types of effects influencing the dynamics behind success for creation and reception processes}. This review explores regularities, mechanisms, and predictive signals in both creation processes (through which individuals, teams, and organizations acquire resources and create new ideas, products, or behaviors, gray boxes) and reception processes (through which a completed product or well-defined behavior is potentially adopted or endorsed by interconnected individuals, teams, and organizations, blue boxes). We identify four interrelated types of effects that influence the collective dynamics behind success for both creation and reception processes: feedback, temporal, network, and identity effects. The boxes highlight key themes found in the literature for each effect type.
 }
        \label{fig:framework2}
    \end{figure}

In the real world, multiple causal factors may influence the collective dynamics behind success. At the same time, individual papers often focus on one or few research questions or hypotheses related to one effect type (e.g., the role of social network structures for career success), which makes it challenging to synthesize the resulting knowledge into one single framework. Therefore, it is helpful to identify the effect types which have received wide and cross-disciplinary attention in the literature.
This review identifies four interrelated types of effects that influence the collective dynamics behind success both in creation and reception processes: feedback, temporal, network, and identity effects (see Fig.~\ref{fig:framework2} for a map). For each type of effect, we review relevant phenomena that concern subjects at one or more of the vertical scales in Fig.~\ref{fig:framework}. While some of these phenomena primarily concern creation processes (e.g., the importance of strong and weak ties for career success), other phenomena are related to reception processes (e.g., the role of influential nodes in the proliferation of a new product in a social network). 

This categorization of different types of effects in Fig.~\ref{fig:framework2} simplifies the navigation of the broad and multifaceted literature on success. It is important to note, however, that the four effect types are interconnected. For example, important identity effects that concern gender differences in personal network structures influence women's opportunities through network effects. Similarly, in creative careers, positive feedback can arise from a central position in relevant social networks through network effects.
Moreover, the creation and reception processes themselves are interconnected, as choices made during creation affect a product’s reception, and the
reception of an individual's products affects the individual’s access to resources for future creation. Consider, for example, the impact of the reception of a scholar’s work on her future funding and employment opportunities. This interconnectedness between effect types and the creation and reception processes points to the complexity of success in socioeconomic systems. We review each of the four effect types below.

\section{Feedback effects}

Here we consider processes in which the subject – whether a product, individual, team, or organization – experiences an initial outcome that serves as a feedback that might increase their likelihood of future successful outcomes. We discuss a widely-studied positive feedback effect (“Success breeds success”), as well as a less intuitive reversal effect (“Failure breeds success”). 

\subsection{Success breeds success}

Many studies on feedback mechanisms are motivated by an interest in explaining the large disparities in success that have been widely observed both for products and their creators. 
Vilfredo Pareto described the disparities he saw in the 19th century with the ``80/20 rule": 80\% of the total wealth was owned by roughly 20\% of the population~\cite{pareto1896course}. 
Since that time, research has documented inequalities in success across diverse domains, including wealth and income~\cite{piketty2014inequality}, scientific citations~\cite{barabasi2012handful,nielsen2021global}, online popularity~\cite{barabasi1999emergence}, and book sales~\cite{sornette2004endogenous}, among others. In many cases, these inequalities have increased steadily over time~\citep{barabasi2012handful,zucman2019global,nielsen2021global}.
Mathematically, this inequality in success is described by a long-tail distribution, meaning that most ideas or individuals experience a low level of success, but a non-negligible fraction of outliers achieves runaway success.

But how does this inequality arise? One possible explanation is that small differences in performance or quality result in large differences in success~\cite{rosen1981economics}. 
Research shows, however, that differences in success do not necessarily reflect differences in performance or quality. Another possible mechanism to explain these observed inequalities is that success, in fact, breeds more success. In 1968, Robert K. Merton suggested this idea, defining the Matthew effect as the process through which highly reputed individuals accrue greater recognition for their work than individuals who have not yet made their mark~\cite{merton1968matthew}. The Matthew effect explains how those with initial, potentially arbitrary advantages may accrue enduring advantages over their competitors through positive feedback, which can create broad inequalities in success and raise questions regarding the extent to which great success is indicative of exceptional ability~\cite{denrell2012top}. 
The Matthew effect--and closely related concepts like cumulative advantage, the rich-get-richer phenomenon, or preferential attachment in network science--has captured sustained attention from diverse disciplines~\cite{merton1968matthew,rigney2010matthew,perc2014matthew}. Individuals connected with more prestigious people or institutions early in their careers may have easier access to more and better resources and rewards~\cite{clauset2015systematic,fraiberger2018quantifying,chowdhary2023dependency}. Their established reputations lead to further recognition, which others may interpret as a proxy for the quality of their work~\cite{podolny1993status}. Hence, early connections have been shown to be strong predictors of success for artists~\cite{fraiberger2018quantifying}, DJs~\cite{janosov2020elites}, and researchers~\cite{sarigol2014predicting,li2019early}, among others.
The Matthew effect also applies to the success of products, which is often accurately predicted by various dimensions of the products' early success, e.g., for online content~\cite{shulman2016predictability}, books~\cite{yucesoy2018success}, scientific papers and patents~\cite{wang2013quantifying,higham2017fame,mariani2016identification,mariani2019early}. 
In addition to the empirical support for the link between early and subsequent success in observational studies, researchers have devised laboratory and field experiments that have demonstrated the causal nature of this link~\cite{salganik2006experimental,salganik2008leading,van2014field}.

\begin{figure*}[t]
        \centering
 \includegraphics[scale=0.75]{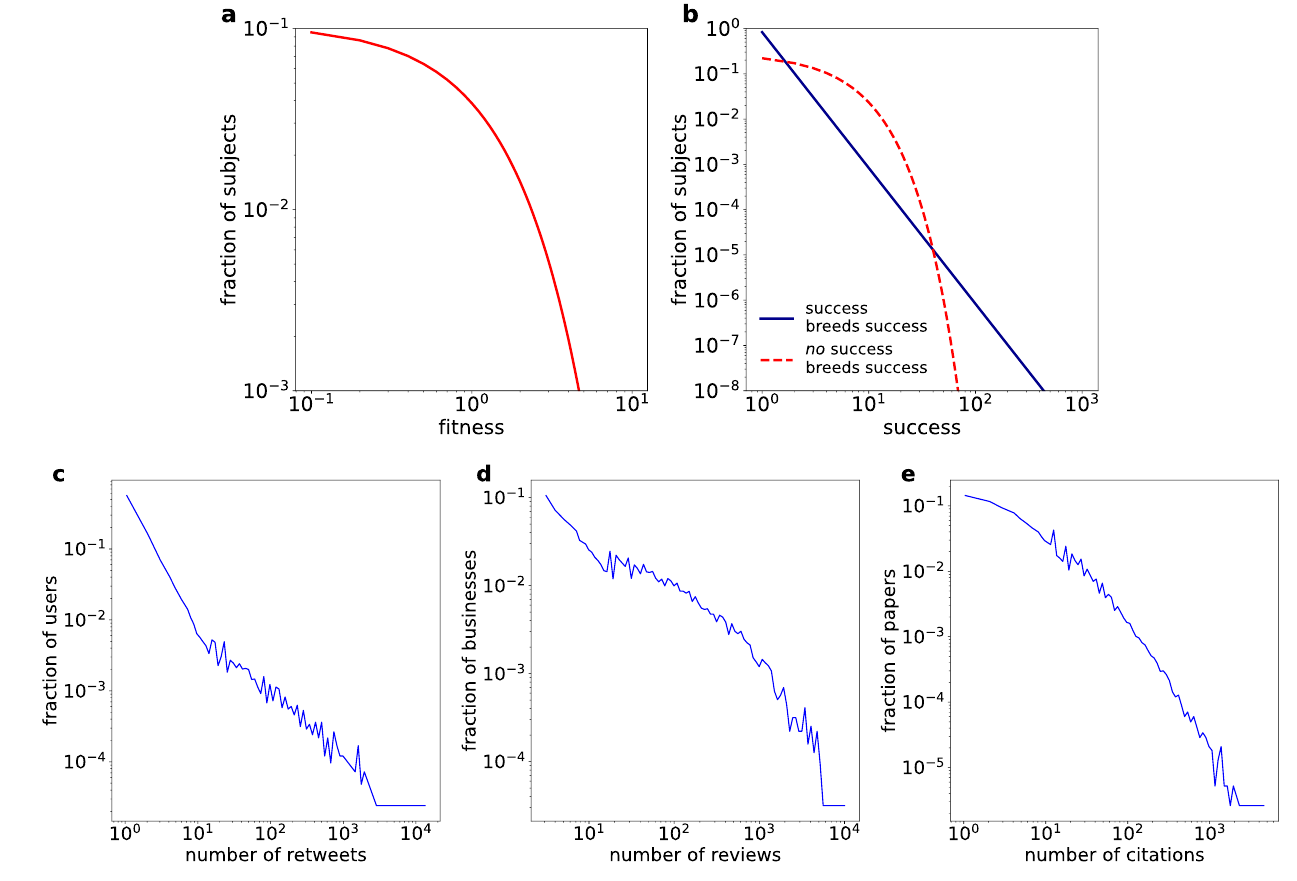}
\caption{\textbf{Fitness and success-breeds-success.} 
The concept of fitness was borrowed from ecology to capture the intrinsic appeal of an idea, product, or behavior. Empirical data suggest that fitness typically follows bounded distributions (e.g., exponential~\cite{kong2008experience,medo2011temporal}, see panel \textbf{a} for an illustration). In a hypothetical world in which recognition is purely determined by fitness, success would also follow a bounded distribution (panel \textbf{b}, red line). However, mechanistic models of success dynamics indicate that even when fitness follows a bounded distribution, success-breeds-success mechanisms can generate fat-tailed success distributions (panel \textbf{b}, blue line) where a few products accrue most endorsements, and many others remain unnoticed. There is ample empirical evidence of fat-tailed success distributions, from wealth and income to scientific citations, from online popularity to book sales, and more (see main text). Here we illustrate: (\textbf{c}) the distribution of and the total number of retweets received by a Twitter user in response to the experimental discovery of the Higgs boson [dataset made publicly available by~\cite{de2013anatomy}]; (\textbf{d}) the number of reviews received by a business in Yelp [dataset made publicly available by~\cite{yelp}]; (\textbf{e}) the number of citations received by a paper published in American Physical Society (APS) journals [dataset made available by the APS~\cite{aps}, version analyzed in~\cite{mariani2016identification}]. }
        \label{fig:fit_vs_succ}
    \end{figure*}

If the success-breeds-success phenomenon were the only driver of the dynamics of success, then the first individuals, teams, or organizations to enter a new field would almost invariably be more successful than latecomers, resulting in an inevitable first–mover advantage~\cite{newman2009first}.
Yet in real-world socioeconomic systems, latecomers often outperform the very first movers in the long run~\cite{bianconi2001competition}.
To explain the success of latecomers (e.g., Google~\cite{barabasi2016network}), researchers borrow the concept of fitness from ecology to capture the intrinsic appeal of an idea or product~\cite{bianconi2001competition,medo2011temporal,wang2013quantifying,higham2017fame,yucesoy2018success,stringer2010statistical}. 
The fitness of a product is typically operationalized as a latent parameter that influences the rate at which the product accrues new adoptions~\cite{bianconi2001competition,kong2008experience}. The roots of a product's fitness are, in general, context-dependent~\cite{barabasi2016network}; For a scientific paper, for example, the fitness parameter could simultaneously capture its relevance, soundness, visibility, and clarity, which jointly determine its appeal to the scientific community~\cite{stringer2010statistical}.
Incorporating the concept of fitness into models of the success-breeds-success dynamic substantially improves the predictability of long-run outcomes, from the long-term impact of scientific  papers~\cite{wang2013quantifying} and patents~\cite{higham2019ex} to book sales~\cite{yucesoy2018success}. 
These fitness-based models also show that a narrow distribution of fitness parameters, i.e., narrow differences in intrinsic fitness or quality, may manifest in a long-tail distribution of success (Fig.~\ref{fig:fit_vs_succ}) or broad success inequalities.
A recent experimental study confirms the importance of fitness, showing that even when high-fitness products were deceptively presented as unpopular to participants, the high-fitness products eventually became more popular than the low-fitness alternatives that were artificially given an initial advantage~\cite{van2019self}. In choice experiments where there is a correct answer, recent evidence indicates that early mistakes made by a group self-correct for easy tasks, but not for difficult ones~\cite{frey2021social}. At the same time, more research is needed to fully clarify under which conditions social feedback effects are strong enough to prevent the highest-fitness products from succeeding~\cite{van2019self}.
The interplay between success-breeds-success, fitness, and the structural factors determining the relative success of products, ideas, and behaviors remains an important domain for future research.

\subsection{Failure breeds success}

Given the impact of early success on later positive outcomes, one might expect that early failures to simply increase the likelihood of future failures. Recent research, however, has challenged this paradigm, revealing surprising benefits of early failures for both individuals and organizations.

There are two simple mechanisms through which early failures could lead to later success.
First, if every attempt has some probability of success, the probability that multiple attempts all fail decreases exponentially with the number of attempts. A individual, team, or organization may, therefore, succeed by chance if enough attempts are made.
Second, individuals, teams, and organizations may eventually succeed if their efficiency increases as they perform a task repeatedly. 
For example, engineer Theodore Paul Wright found in 1936 that the labor cost of producing an airplane decreases as a function of the total number of airplanes produced, following a power law~\cite{wright1936factors}. In general terms, the regularity he found (often referred to as Wright's law) suggests that the time needed for an organization to formulate the $n$-th attempt, $t_n$, decreases with the number of previous attempts as a power-law: $t_n/t_1\sim n^{-\gamma}$, where $\gamma$ is the learning rate. 
Since Wright's finding, this law has been used to empirically measure performance improvements in diverse manufacturing processes~\cite{dutton1984treating}. 

A recent study~\cite{yin2019quantifying} highlights the importance of learning from failures by examining successful and unsuccessful attempts across different fields, including grant applications submitted to the National Institutes of Health (NIH), startups seeking an initial public offering (IPO), and security settings. In each case, the penultimate attempt before a success are notably better than the first, indicating that improvement, not luck, is often at play, which is consistent with the learning theory.  

\begin{figure*}[t]
        \centering
 \includegraphics[scale=0.35]{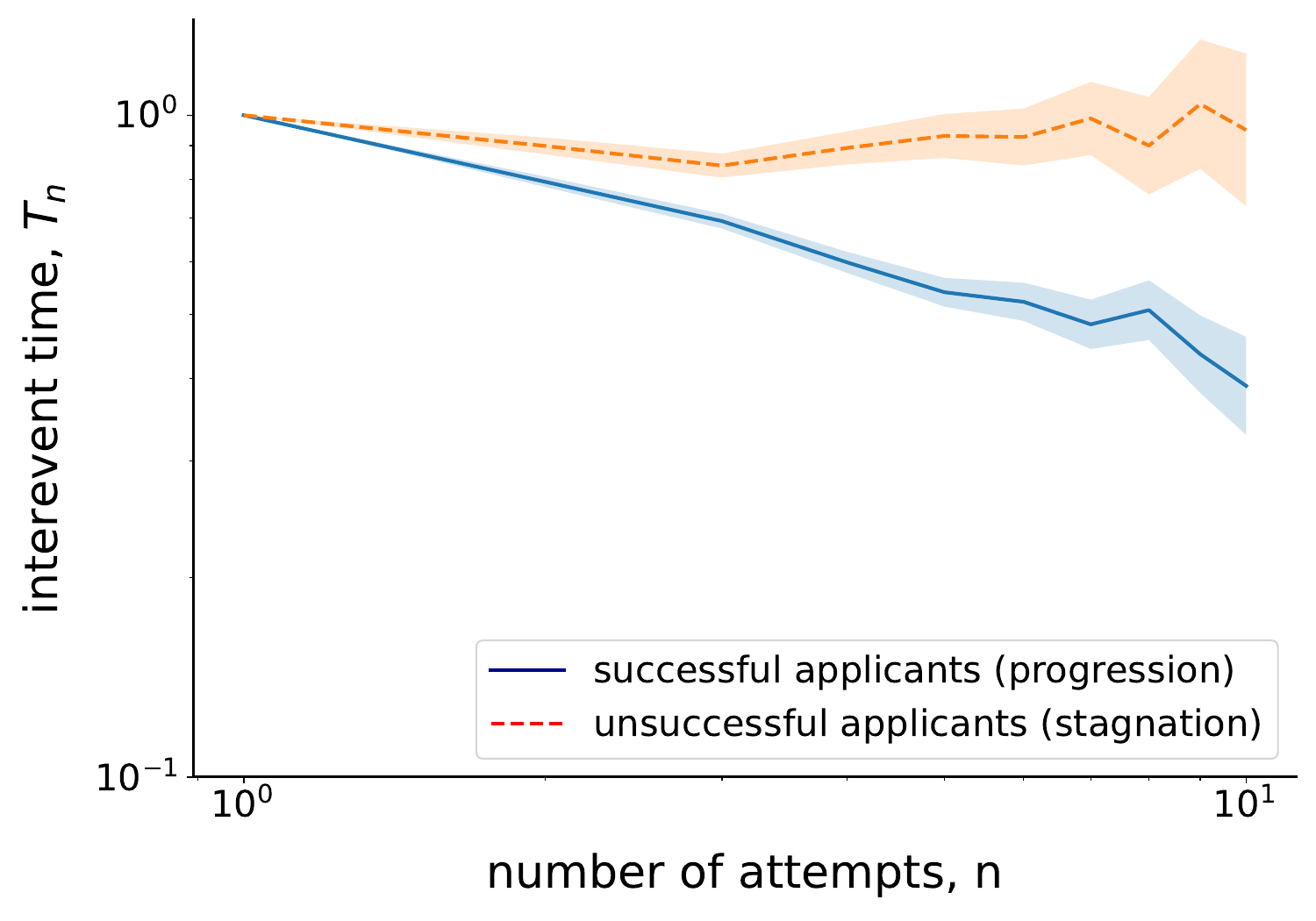}
\caption{\textbf{Early signals of the failure-to-success transition.} For successful grant applicants in the NIH data (blue line), the average inter-event time between two consecutive failures, $T_n$, is a decreasing function of the number of early failures, $n$, as predicted by Wright's law. This means that it takes less time for the applicant to formulate a new attempt, i.e., the applicant is in the progression regime. Unsuccessful applicants, however, do not follow Wright's law (orange line).
The significant difference between the two lines can be leveraged to make accurate success vs. failure predictions in science and entrepreneurship.
Data from~\cite{yin2019quantifying}.}
        \label{fig:failure}
    \end{figure*}

However, the learning theory is contested by two major observations. 
First, Wright’s law only holds for individuals who eventually succeed. The performance and efficiency of those who do not succeed, which includes a substantial portion of the population, do not improve over time (see Fig.~\ref{fig:failure}). Second, failure streaks are longer than expected, following a fat-tailed distribution, which suggests that individuals may get stuck in suboptimal strategies. These surprising empirical observations can be captured by a one-parameter model in which each new attempt aims to improve over previous efforts by generating new components or by borrowing the best-performing components from previous attempts. The model is amenable to an analytical solution that unveils a phase transition between a regime of progression, in which performance improves over time according to Wright’s law, and a regime of stagnation, in which individuals do not learn significantly from past mistakes and, consequently, their performance score experiences saturation~\cite{yin2019quantifying}.
This model allows for accurate out-of-sample success vs. failure predictions by analyzing only the distribution of individuals' inter-event times in the early phase of their careers~\cite{yin2019quantifying}.

Despite these findings, our understanding of failure and its relationship with success remains limited. 
Researchers have begun to consider the causal link between early setbacks and future career outcomes. A recent study finds, for example, that junior NIH grant applicants whose proposals fell just below an arbitrary funding cutoff outperformed those whose proposals fell just above it in the long-run~\cite{wang2019early}. 
 One potential explanation of these findings is a screening mechanism, where early-stage failure may screen out less determined individuals. 
But, evidence suggests that the performance boost associated with failure appears to go beyond this explanation~\cite{wang2019early}, echoing Friedrich Nietzsche's adage, ``what does not kill me makes me stronger"~\cite{nietzsche1889twilight}. 
Empirically, these findings are consistent with patterns observed in  resubmission of rejected manuscripts~\cite{calcagno2012flows} 
as well as psychology research that shows 
that failure may teach valuable lessons that are hard to learn otherwise~\cite{sitkin1992learning} or may motivate individuals to increase their efforts in future endeavors~\cite{burnette2013mind}.

Overall, these findings substantially inform the ensuing debates about whether great success truly reflects great potential~\cite{merton1968matthew,salganik2006experimental,van2014field,bol2018matthew,gladwell2008outliers,barabasi2018formula,fraiberger2018quantifying,rosen1981economics,van2019self}. 
Indeed, one of the most consistent findings in social and behavioral sciences is that past success is a strong predictor of future success (See Section Success breeds success). 
While these success-breeds-success dynamics may channel resources to high performers, they also typically foster ``winner-take-all" systems, exacerbating inequities.
Understanding the link between failure and achievements therefore has profound implications not only for resource allocation but also  for the nurturing and broadening of potential talent pools.

\section{Temporal effects}

This section reviews patterns and mechanisms associated with variations in success over time, which include temporal changes in the reception of a product, as well as changes in the creation and reception of cultural products over a creator's career.

\subsection{Aging of ideas and products}

Time plays an important role in the success of both products and careers. How does success change over time? How much success will a product or an idea accrue throughout its lifetime? Can we achieve accurate early-stage predictions from initial signals? 
The answers to these questions are relevant to diverse decision-makers, as early-stage success predictions for new products and services often inform resource allocation decisions~\cite{golder1997will}. For example, early detection of significant technologies is key for firms’ and investors' R\&D investment decisions~\cite{kogan2017technological}, and early detection of viral content in online domains is critical to preventing the diffusion of misinformation~\cite{bak2022combining}.

There is a long history of studying temporal patterns in the collective adoption of new products and technologies in the innovation diffusion literature~\cite{bass1969new,rogers2010diffusion}. 
The Bass model~\cite{bass1969new}, for example, describes changes in the success of new products over time and shows that 
the adoption of a new product tends to start slow, increase rapidly, and eventually reach saturation as the entire market is captured, resulting in an S-shaped diffusion curve~\cite{bass1969new,rogers2010diffusion}. While this seminal model offers a parsimonious description of diffusion curves, it has unrealistic features, which has motivated researchers to leverage large-scale, time-stamped data to characterize temporal patterns in the adoption of new products more accurately. 

Recent research focuses on how the success of ideas and products decays over time, departing from the Bass model's framework~\cite{medo2011temporal,wang2013quantifying,higham2017fame,yucesoy2018success,candia2019universal}. 
These studies cover the temporal decay of success for a remarkably wide range of products, including tweets and memes on Twitter and similar social media platforms~\cite{leskovec2009meme}, YouTube videos~\cite{crane2008robust}, books~\cite{sornette2004endogenous,yucesoy2018success}, scientific papers and patents~\cite{valverde2007topology,wang2013quantifying, medo2011temporal}, and new technologies~\cite{weiss2014adoption}. 
Most of these works focus on empirical observation and modeling the aging effect~\cite{valverde2007topology,medo2011temporal,wang2013quantifying}, defined as the tendency of older items to accrue new adoptions at a slower rate than newer items. Conclusions regarding the functional shape that best captures the aging effect vary by context,
but the temporal functions that describe how success decays generally converge in the long run, suggesting that it is possible to predict a product’s ultimate level of success~\cite{stringer2010statistical,wang2013quantifying,yucesoy2018success}.

Within almost every domain, however, there are notable exceptions that defy aging effects and exhibit distinct temporal dynamics. In science and technology, for example, there are ``sleeping beauties" --papers~\cite{van2004sleeping} or patents~\cite{hou2019patent} -- that experience a new surge in citations after a long hibernation period of weak or moderate attention. These awakenings could be due to interest from a different research community or, in the case of application-oriented papers, the first citations by patents~\cite{van2017sleeping}. Analyses of millions of manuscripts indicate that sleeping beauties are widespread across the sciences, as revealed both by citation data~\cite{ke2015defining} and online attention received by papers~\cite{zakhlebin2020diffusion}.
Exogenous influences may also affect the temporal patterns of success, causing bursts and sharp peaks of interest in specific products, as shown in online domains~\cite{crane2008robust}, book sales~\cite{sornette2004endogenous} and academia~\cite{mazloumian2011citation}. 

Like fitness, temporal effects counterbalance older products' advantages due to success-breeds-success dynamics. Researchers have developed flexible models that integrate aging effects with these fitness and the success-breeds-success dynamics, applying them to describe the full success lifecycle of products across different domains, ranging from marketing~\cite{rogers2010diffusion} to science~\cite{wang2021science} and cultural domains~\cite{yucesoy2018success}. 
These models enable accurate long-term predictions based on early success trajectories~\cite{wang2013quantifying,yucesoy2018success}, but this concept remains largely underexplored, highlighting an interesting direction for future research.

\subsection{Career lifecycles}

Temporal effects matter not only for the success of ideas and products, but also for the success of individual careers. One of the primary questions related to career success asks when individuals do their greatest work. 
This question has important implications for institutions that aim to select and nurture talent. 
In academia, for example, early-stage researchers who have not yet produced their highest-impact work may be underestimated, which could potentially lead to less funding or discourage them from continuing on the academic career path. 

While the timing of career success is necessarily domain dependent, research reveals reproducible patterns across a diverse set of careers. It is commonly held that young minds are most likely to produce groundbreaking work. 
In fact, Einstein once remarked, ``A person who has not made his great contribution to science before the age of thirty will never do so''~\cite{barabasi2018formula}.
However, younger individuals may lack elements that facilitate career success, including specialized knowledge that requires time to acquire~\cite{jones2009burden}, strong reputations~\cite{petersen2014reputation}, social capital~\cite{li2019early}, and experience~\cite{sekara2018chaperone}.
The economics literature has consistently reported an inverted U-shaped dependence of an individual's success on their age. Although the precise position of the curve’s peak depends on both context and historical periods~\citep{jones2011age}, its presence is ubiquitous across diverse domains~\cite{jones2011age, azoulay2020age}.

A pessimistic interpretation of this curve is that individuals become less likely to produce groundbreaking work as they age, which is analogous to the aging effects in new product diffusion described above.
However, recent observational results challenge this perspective by demonstrating that any piece of work in a given individual's creative career--from scientific papers~\cite{sinatra2016quantifying} to movies, books, and music~\cite{liu2018hot,janosov2020success}--has the same likelihood of becoming their most impactful work. 
Dubbed the random impact rule~\cite{sinatra2016quantifying}, this apparent lack of regularity provides empirical support to previous theories of creativity and innovation, such as Simonton’s idea~\cite{simonton2004creativity} that creative outputs throughout a career have equal odds of success. 
This observation can be reconciled with the widely-observed inverted U-shaped relation between individual success and age by noting that as individuals age, they tend to become both less successful and less productive. Productivity--rather than age--may, therefore, drive success: A skilled individual's likelihood to create a hit may remain high as they age as long as they keep producing~\cite{barabasi2018formula,wang2021science}. 

Research also indicates that productivity does not entirely explain the success of individuals' work. A recent study finds that an individual's highest-impact works in science, the movie industry, art~\cite{liu2018hot}, and social media~\cite{garimella2019hot} are often clustered in close succession, occuring in a ``hot streak". Reminiscent of the controversial idea of ``hot hands" in professional sports~\cite{bar2006twenty}, these bursts of successful creative activity tend to be uniformly distributed during a career, and they are not characterized by increased productivity~\cite{liu2018hot}, consistent with the random impact rule. Yet, hot streaks can be recognized at their onset. In show business, a supervised machine learning algorithm can capture signs anticipating an actor's annus mirabilis--their best career year--with up to 85\% accuracy~\cite{williams2019quantifying}.

The ubiquity of the hot streak phenomenon across diverse professions has also inspired research to determine the mechanisms that drive it. Recent findings on scientific and artistic careers~\cite{liu2021understanding} indicate that the onset of hot streaks is associated with detectable patterns of exploration and exploitation. Before a hot streak, individuals tend to explore a wide range of topics. Then, at the onset of their hot streak, they focus on a restricted set of topics and styles~\cite{liu2021understanding}. Moreover, recent theory hypothesizes that both the timing and the sustainability of creators' success may be influenced by their early exploration patterns~\cite{berg2022one}. 
In particular, the novelty and variety of the products creators create before their first big hit may positively influence their likelihood of achieving subsequent hits. High novelty in a creator’s early works, though, might be associated with a lower likelihood of generating a first big hit, which suggests an early-stage risk-return tradeoff associated with novelty. Empirical evidence on US singers confirmed these hypotheses~\cite{berg2022one}, which might be further tested and refined by future studies in other domains.

\section{Network effects}

The temporal evolution of both career success and new product diffusion necessarily depends on the underlying structure of relevant social and economic networks. Network effects operate in both creation and reception processes. In reception processes, the structure of the relevant social networks of potential adopters affects the large-scale adoption and success of new ideas, products, and behaviors (see Section: Simple vs. complex contagion).
In creation processes, the structure of the social networks around an individual creator affects the success of her career (see Section: Social ties that influence career success), and the internal structure of a team influences its success (`see Section: Team structure and success).

\subsection{Simple vs. complex contagion}

\begin{figure*}[t]
        \centering
        \includegraphics[scale=0.75]{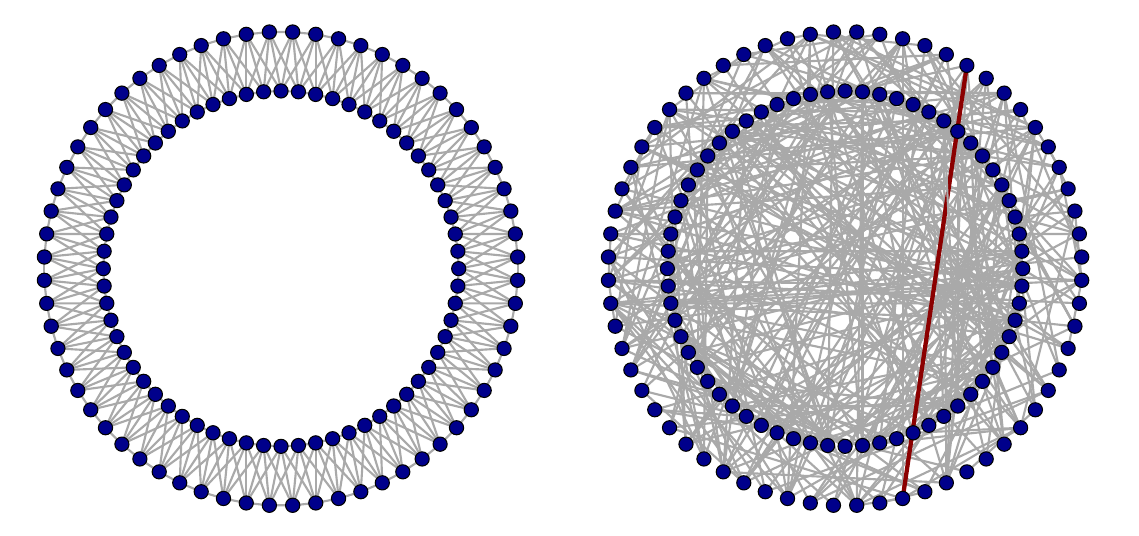}  
\caption{\textbf{Which network structure favors the large-scale adoption of a new product or behavior?} 
The right-side, low-diameter network features long ties that connect otherwise distant nodes (e.g., the red tie). The presence of more long ties is associated with a lower diameter (i.e., a smaller average distance between the nodes, which implies the small-world property~\cite{barabasi2016network}) and lower clustering (i.e., lower tendency of an individual's social contacts to be connected). By contrast, long ties are absent in the left-side, high-diameter network. Our intuition and the properties of epidemic spreading processes~\cite{watts1998collective} suggest that an idea, product, or behavior would spread faster and farther in the right-side network.
While this intuition is correct for simple social contagions, it is not supported for complex social contagions, i.e., adoption processes in which multiple exposures to a new behavior or product are required before an individual adopts. Complex contagions, such as the adoption of health-related behaviors, spread more effectively in the left-side network, as demonstrated both theoretically~\cite{centola2007complex,guilbeault2021topological} and experimentally~\cite{centola2010spread}.
Data from [Centola, D. The spread of behavior in an online social network experiment. \textit{Science} 329, 1194–1197
(2010). DOI: \url{10.1126/science.1185231}].
}
        \label{fig:nets}
    \end{figure*}

Success-breeds-success, fitness, and aging effects are powerful concepts for modeling and predicting the aggregate dynamics of success across diverse domains. 
The literature indicates, however, that to engineer the success of new products and behaviors, interpreted here as achieving their large-scale adoption, modeling and predicting the aggregate dynamics are insufficient. Social networks of interconnected potential adopters also have a significant influence on the adoption process due to various social contagion mechanisms that affect how information flows in social networks and how an individual's adoption choices influence peers' choices, including network externalities, social learning, normative influence, and direct interpersonal communication or word-of-mouth (see~\cite{peres2010innovation,dimaggio2012network,muller2019effect} for comprehensive reviews). 
These social contagion mechanisms imply that the success dynamics of a new product or behavior can be described as a spreading process through the social networks of potential adopters.
This description motivates multiple policy-relevant questions, including: Which network structures facilitate or impede the spread of a new product, idea, or behavior? In a given social network, which nodes should be targeted first to initiate the spread of a new product or behavior? How can a fast-growing spreading process, such as the large-scale spread of dangerous misinformation, be slowed?

Many observers draw analogies between the spread of viruses and the spread of ideas, products, and behaviors through social contacts.
This analogy to pathogen contagion, however, has been heavily contested in the literature. 
The key insight lies in the fundamental distinction between simple and complex contagions~\cite{centola2021change,guilbeault2018complex}. Biological viruses spread through simple contagion in which a single exposure to an agent can be sufficient for transmission to occur~\cite{barabasi2016network}. This mechanism also applies to the spread of information (e.g., news, Twitter hashtags, and memes) concerning conversational topics~\cite{notarmuzi2022universality}, online misinformation, and fake news~\cite{del2016spreading,vosoughi2018spread}. 
By contrast, complex contagions require social reinforcement: an individual must have multiple exposures to an idea or a behavior to adopt it~\cite{centola2007complex}. 
Complex contagion more accurately describes spreading processes that require individuals to make a substantial personal investment due to the costs or risks involved, including reputational or social risks (e.g., the risk of being alienated by a social circle for adopting a political position), personal risks (e.g., in social protests), and personal effort~\cite{guilbeault2018complex,centola2021change}. 
Empirical evidence of complex contagion has been found for the adoption of health behaviors, online social platforms, new technologies, tweets and hashtags related to political views and controversial political behavior, and social movements and political protests -- see \cite{guilbeault2018complex} for an extensive review.

The distinction between simple and complex contagions helps us understand which network structures may facilitate or impede the large-scale adoption of a new idea, product, or behavior (Fig.~\ref{fig:nets}).
The spread of simple contagions is highly effective in small-world networks where long ("structurally weak") ties reduce the average distance between pairs of nodes~\cite{watts1998collective}.
By contrast, the spread of complex contagions is more effective in highly clustered networks where a node's social contacts tend to be connected as well, even if the network has a larger diameter and contains fewer long ties. 
Low clustering and long ties, therefore, are only conducive to success for simple contagions; they can stifle the spread of complex contagions~\cite{centola2007complex,centola2010spread}
because clustering is needed to form bridges that are wide enough to enable the necessary social reinforcement~\cite{centola2007complex}.

This distinction is also relevant to the  ``influence maximization" or ``seeding" problem, i.e., the choice of 
the best subsets of network members with whom to initiate the spread of a new idea, product, or behavior in order to maximize the likelihood of large-scale adoption~\cite{muller2019effect}. 
The idea that a few influentials may have a disproportionate effect on spread has inspired a series of studies on social hubs, trendsetters, influencers, and influence maximization in computer science~\cite{domingos2001mining}, management science~\cite{hinz2011seeding,muller2019effect}, and network science~\cite{lu2016vital}, among others.
Across these diverse domains, researchers have developed various network-based algorithms to identify effective influencers and, in some cases, validate them via simulations under different scenarios~\cite{lu2016vital} and field experiments~\cite{hinz2011seeding,paluck2016changing,airoldi2024induction}.
Sometimes, however, greater spread is instead achieved by building a critical mass of easily influenced people rather than by targeting a few central influencers~\cite{watts2007influentials,gonzalez2011dynamics,lanz2019climb}.
Recent theory indicates that for complex contagions, a critical mass can be built not by targeting the social hubs, but by targeting network locations that are traversed by wide enough bridges, which provide the level of social reinforcement needed for the transmission of complex contagions~\cite{guilbeault2021topological}.

Overall, the concept of ideas and behaviors spreading throughout social network structures is incredibly nuanced. The dichotomy between simple and complex contagion demonstrates that a network topology or influence maximization strategy that facilitates the success of simple contagion processes can doom complex contagion processes to failure~\cite{centola2021change}.
The interdependence of network structure, contagion type, and diffusion success makes the design of effective interventions to maximize or prevent spread a formidable challenge. The persistence of misinformation and the difficulties faced by public health policies during the COVID-19 pandemic serve as a prime example of this challenge. 
Future research may increasingly focus on network experiments, which could systematically validate theoretical predictions of spread patterns conditional on network structure and the underlying contagion mechanism.

\subsection{Social ties that influence career success}

Social networks are relevant not only for the spread of new products and behaviors, but also for the success of individual careers.
This long-standing realization has motivated several interrelated questions: Which network positions and social ties predict and influence the most career success? For a given individual, what are advantages and disadvantages of different network positions? In what circumstances are certain network positions preferable?

\begin{figure*}[t]
        \centering
 \includegraphics[scale=0.65]{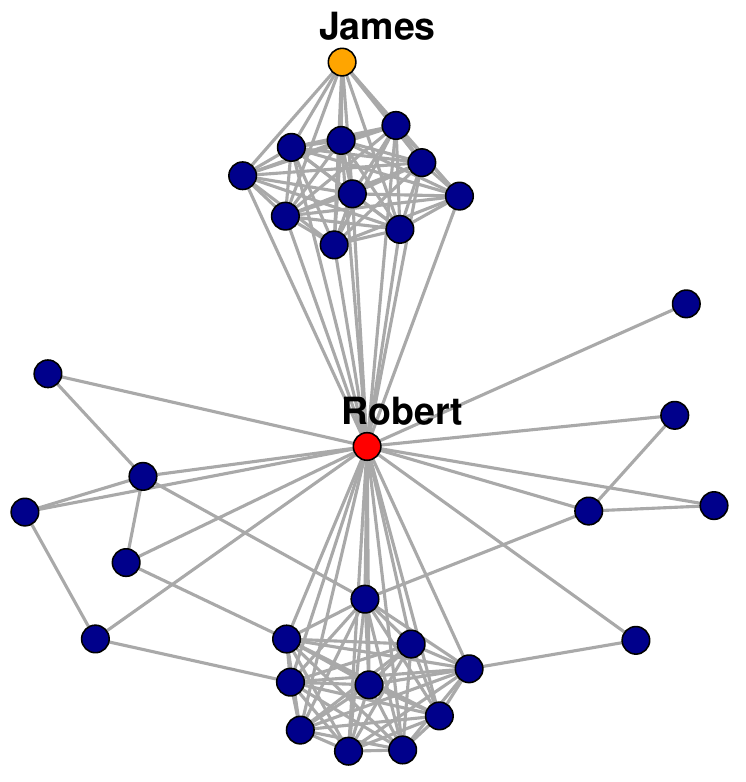}
\caption{\textbf{Which network position favors individual success?} Consider the illustrative social network structure depicted here, where the nodes represent individuals and the gray lines represent their social ties. According to Burt's structural hole theory~\cite{burt2018structural}, Robert (red node) is positioned as a broker, as he connects otherwise disconnected network communities through structurally weaker ties (i.e., ties to people with whom he shares relatively few social contacts). James' ties (orange node) are structurally stronger and embedded within a cohesive community. Recent works have leveraged large-scale datasets and digital experiments to understand the circumstances in which weak or strong ties are more beneficial to the individuals' information advantages and economic outcomes, and to measure the causal effects of tie strengths on successful job transmission (see main text). } 
        \label{fig:advantage}
    \end{figure*}

Pioneering work by Granovetter~\cite{granovetter1973strength} shows that people are more likely to find higher-level managerial jobs through personal contacts than through advertisements or job applications. And surprisingly, the contacts most useful for finding a job are not one’s strong ties -- close friends or family. Rather, they are acquaintances -- the more distant relationships known as weak ties. While strong ties tend to have similar knowledge bases and thought patterns, weak ties can provide access to novel information by bridging otherwise disconnected social spheres. 
Burt’s structural hole theory is tightly related to this idea. It posits that ``broker” or ``bridge" individuals may benefit from occupying structural holes to connect otherwise disconnected communities~\cite{burt2004structural} (see Fig.~\ref{fig:advantage}). This position allows them to potentially provide these communities with more diverse and locally-scarce information, placing them in a unique position to recombine existing knowledge and develop innovative ideas~\cite{burt2004structural,burt2018structural}.

Managing the opportunities provided by weak ties and bridging positions, however, comes at a cost. Because of the additional effort needed to bridge disparate groups, serving as a broker can be a risky and time-consuming endeavor~\cite{sun2021interdisciplinary}.
Moreover, weak ties are not necessarily more effective than strong ties in achieving information advantages and successful outcomes. For example, complex knowledge is often transmitted more effectively through strong ties~\cite{hansen1999search}, e.g., through mentor-protégé ties~\cite{podolny1997resources,malmgren2010role}; 
stronger ties tend to carry more influence than weaker ties~\cite{coleman1988social,reagans2003network}; and ties of intermediate strength may be most effective ones for information transfer~\cite{onnela2007structure}.
Strong ties embedded in a cohesive network, then, may also lead to the success of individuals, teams, and organizations~\cite{coleman1988social,uzzi1996sources}.

Recent studies~\cite{aral2011diversity,aral2022exactly,rajkumar2022causal} have explored the conditions under which we can expect weak vs. strong ties to offer individuals the greatest benefit in terms of information advantages and economic outcomes, such as project completion~\cite{aral2011diversity}, landing new jobs~\cite{rajkumar2022causal}, and achieving economic prosperity~\cite{jahani2023long}. 
This work finds that information from structurally weak (or long) ties--those that connect people who share few mutual social contacts--is more diverse and locally-scarce~\cite{aral2022exactly} than information from structurally strong ties. This information, though, also arrives at slower rates, as the frequency and volume of interactions with weak ties tend to be lower~\cite{aral2011diversity} (see \cite{jahani2023long} for exceptions).
This ``diversity-bandwidth tradeoff"~\cite{aral2011diversity} raises questions concerning the novelty of the information that strong and weak ties deliver over time. 
Recent theory and observational findings suggest that the answer depends on properties of the information ecosystem in which the individual operates. 
Weak ties are more beneficial in environments with fewer topics, slowly-changing information, and high heterogeneity in the information individuals possess~\cite{aral2011diversity}.
Here, the definition of novelty matters, as well. Strong ties deliver more novel information in terms of diversity and non-redundancy, but weak ties tend to deliver more unique information (i.e., information that is more distinct from that delivered by other social contacts)~\cite{aral2022exactly}.
In terms of economic outcomes, recent population-scale observational evidence indicates that individuals who exhibit structurally weak ties with high bandwidth tend to have higher levels of economic prosperity~\cite{jahani2023long}.

While these studies rely on observational data, a recent experimental study \cite{rajkumar2022causal} provides the first causal link between tie strength and an individual-level economic outcome: successful job transmission. Transmission is assumed to occur when a user reports working for a company where her social contact (established more than a year before) has already reported working for more than a year. Multiple large-scale randomized experiments on LinkedIn reveal that moderately weak ties--not the structurally weakest or the strongest ties--lead to the highest job transmission probability~\cite{rajkumar2022causal}.
We expect future research to include more experimental studies to systematically validate theoretical predictions regarding the relative advantages of weak vs. strong ties and other characteristics of an individual's network for information advantages as well as various performance and labor market outcomes.

\subsection{Team structure and success}

Network structures affect not only individuals’ career success, but also the whole network’s success. Research on team success illustrates this point particularly well.
A team can be viewed as a network of individuals that achieves certain performance outcomes through a group interaction process~\cite{hackman1975group}. 
In the wisdom of the crowd literature, for example, participants team up to solve estimation tasks~\cite{centola2022network}; in science and R\&D teams, researchers team up to solve scientific and technological problems~\cite{wang2021science}. The resulting collective dynamics of interactions can lead to the emergence of a collective intelligence -- as first proposed by Spearman at the beginning of the 20th century~\cite{spearman1904general} -- which does not correlate with the average or maximum intelligence test scores of the team's individual members, but is associated with reciprocity, social sensitivity, and the presence of women~\cite{woolley2010evidence}.
Importantly, viewing the team as a network of interacting members suggests that structural features of the interactions network could predict team performance, which is observed in contexts as diverse as sports~\cite{wasche2017social}, task solving~\cite{leavitt1951some}, and management~\cite{hansen2001so}.

Is there an optimal team network structure? In the early 40s, seminal experiments by Leavitt~\cite{leavitt1951some} provided evidence that centralized network structures outperform decentralized structures in task solving.
Yet, subsequent studies showed that for success dimensions related to complex problems and environments (e.g., financial trading performance and scientific innovation), decentralized networks tend to outperform centralized networks~\cite{shaw1964communication,saavedra2011synchronicity,brackbill2020impact,centola2021change,centola2022network}.
Centralized structures favor rapid, top-down communication and can be highly effective for solving simple tasks~\cite{centola2022network}.
But for solving complex tasks, such informationally-efficient networks tend to converge rapidly on good solutions but not optimal ones. In informationally-inefficient networks (e.g., high-diameter, high–clustering networks, see Fig.–\ref{fig:nets}), the team members have lesser exposure to the ideas and solutions developed by others. Consequently, they explore the solution space more widely and, as a result, are more likely to discover optimal solutions, as demonstrated both theoretically~\cite{lazer2007network} and, more recently, experimentally~\cite{brackbill2020impact}.
Similarly, in collaborative creative environments, informationally-efficient small-world structures can be beneficial to promote creative flows between individuals and positively affect financial and artistic success in collaborative tasks, but only up to a threshold beyond which the effect is reversed~\cite{uzzi2005collaboration}.

Interactions’ specifics and characteristics can also affect team performance. For example, a recent experimental study on collective estimation tasks~\cite{almaatouq2020adaptive} found that two adaptive interaction mechanisms, namely network plasticity (i.e., whether a given team's members are allowed to alter their connections) and feedback (i.e., whether the participants receive feedback on the accuracy of their own and others' estimates) can improve team performance.
The emotional load of interactions can also play a role, as revealed by a recent study on teams in escape rooms~\cite{o2022anatomy}. 
The relative importance of teams’ structural and interaction-level properties for team success is a promising direction for future research, which could benefit not only from traditional network analysis~\cite{wasche2017social,centola2022network}, but also from novel methods to characterize higher-order network interactions~\cite{battiston2020networks}.

\section{Identity effects}

At the heart of most research on success lies the sobering recognition that our societies present many departures from meritocracy. Identity and demographic characteristics impact success, and this effect can, in turn, influence outcomes at the scale of groups and institutions. The literature on the link between success and identity faces the puzzling dilemma that diversity is essential for solutions to the complex problems that knowledge-based economies face~\cite{page2017diversity,hong2001problem,hong2004groups,hoogendoorn2013impact,yang2022gender}, and yet identity effects are typically taxing for under-represented individuals. For example, while experiments have found that the presence of women is associated with increased group intelligence~\cite{woolley2010evidence} and gender diversity is linked to improved team performance~\cite{hoogendoorn2013impact}, women still face a pay gap~\cite{aragao2023gender,goldin2017expanding}. Moreover, a longitudinal study in Britain, Germany and Switzerland has suggested that this disadvantage in earnings can trickle down to penalise entire domains where women are in the majority~\cite{murphy2016feminization}. A recent survey of US professionals in science, technology, engineering, and mathematics (STEM) indicates that several intersectional gender, race/ethnicity, sexual identity, and disability status categories are often further disadvantaged, achieving less social inclusion, fewer career opportunities, and reduced rewards than both heterosexual men and women without disabilities~\cite{cech2022intersectional}. Substantially more work is, therefore, needed to uncover the impact of various aspects of identity such as race and ethnicity~\cite{levine2014ethnic,silbiger2019unprofessional,kozlowski2022intersectional}, sexual orientation~\cite{birkett2014sexual}, and disabilities~\cite{lockhart2024authors} on success. 

A new body of research investigates the role of gender in success, making this aspect of identity arguably the best understood to date and a cornerstone in providing methodologies for studying identity effects more broadly.
Research shows that women receive less recognition for their achievements~\cite{sarsons2017recognition,ross2022women}, which may further penalize them, especially during hiring processes that may already be characterized by a subtle gender bias~\cite{moss2012science}. Women scholars have been shown to receive fewer awards, such as Nobel Prizes~\cite{lunnemann2019gender}. When women do receive an award, the prize may be associated with less money and prestige, as has been documented in the biomedical domain \cite{ma2019women}. These large-scale empirical findings complement seminal work on the so-called ``Matilda effect", which describes a hesitance to acknowledge the achievements of women scientists and a tendency to credit men instead~\cite{rossiter1993matthew}. Women employees~\cite{ibarra2010men} and scholars~\cite{regner2019committees} are less likely to be promoted than men, often facing a glass ceiling in success. These disparities hold in traditionally male-dominated areas, such as STEM fields and upper management, as well as in areas with relatively better female representation, such as physiology, language and behavioral sciences~\cite{regner2019committees}. 
Inequalities in recognition persist in online spaces as well -- e.g., in the online dissemination of scientific work~\cite{vasarhelyi2021gender,song2024can} -- indicating that new platforms and intelligent technologies will not level the playing field without additional system-level interventions.

Complex environments, imbalanced representation, and the need for unique success strategies are likely reasons why women perceive failure more negatively, as found by a survey study with academics~\cite{silbiger2019unprofessional}. This study shows that scientists across four intersecting categories of gender and race/ethnicity receive unprofessional peer review comments equally often. However, traditionally underrepresented groups in STEM fields are most likely to perceive the negative impacts of such feedback on scientific aptitude, productivity, and career advancement~\cite{silbiger2019unprofessional}. While this research focused on the perception of negative feedback, other works investigated women's perception of negative stereotypes. Beliefs that raw talent is required for success can make women vulnerable to negative stereotypes, which could explain why academic fields believed to require brilliance have lower female representation~\cite{leslie2015expectations}. More alarmingly, stereotypes regarding men's brilliance are familiar even to six- and seven-year-olds~\cite{bian2017gender}, indicating that misguided ideas about the intellectual abilities of men and women may hold back girls' aspirations and future success.

Some gender inequalities, such as the gap in productivity, disappear once career length is taken into account~\cite{zeng2016differences,jadidi2018gender,huang2020historical}. Women are typically active in the workforce for shorter periods than men, which explains their lower output. However, these shorter careers are, in many cases, not driven by personal choice. Women age out of the movie and music industries, for example, faster than men~\cite{williams2019quantifying,wang2019gender}. This insight highlights the importance of understanding the factors that could better embed women in the workforce and position them for long-term success.

Differences in productivity and prominence can also be explained by collaborations~\cite{li2022untangling}, making professional networks key for success. We have described above how networks can help individuals access new information and resources, which could improve career possibilities. 
However, women face unique challenges in networked labor markets. Empirical evidence from the US film industry indicates that when women work in dense, cohesive networks, they have lower chances of career success than men~\cite{lutter2015do}. The same study also shows that this gender disadvantage is not present when women are allowed to cooperate, throughout their career, in more open and diverse networks.
A possible reason is that in cohesive networks, women are more likely to be constrained by gender-homophilous ties (i.e., connections with other women)~\cite{lutter2015do}. This homophily is typical both in traditional organizational networks and platform-mediated collaborations~\cite{mcpherson1982women,ibarra1997paving,horvat2017gender}. However, induced homophily (due to structural or institutional segregation) and choice homophily (due to personal preferences), are associated with different success outcomes for women~\cite{mcpherson1987homophily}. Since women typically hold lower-status positions and have less access to key decision-makers, induced gender homophily can particularly disadvantage them~\cite{brass1985mens,lutter2015do}. At the same time, choice homophily can be helpful in providing gender-related tacit knowledge as long as it is supplemented by diverse weak ties spreading job market information~\cite{yang2019network}. 

Given these intricacies in group dynamics, women leverage their team memberships differently. 
Gender-balanced teams often outperform gender-homogeneous ones, as shown by field experiments on management tasks~\cite{hoogendoorn2013impact}, observational data from academia~\cite{yang2022gender}, and laboratory experiments on diverse tasks~\cite{woolley2010evidence}. Proposed reasons why the presence of women can improve team performance include their greater social sensitivity, interpersonal orientation, and egalitarian behaviors, which together improve the team's collective intelligence~\cite{woolley2010evidence,bear2011role}.
Notwithstanding this evidence, extensive management literature documents difficulties with leveraging gender-diversity in teams when in-group bias exists~\cite{shore2009diversity}.
Additionally, biased self-assessment may influence the likelihood of success.
For example, a survey study in a governmental agency found that gender and racially diverse teams rated their performance less favorably than homogeneous teams, even when ratings by external examiners showed no significant difference~\cite{baugh1997effects}.
Overall, empirical evidence indicates that the effect of gender diversity on team performance depends on both the organizational context and personal perceptions~\cite{baugh1997effects,bear2011role}, which points to the importance of building inclusive work environments~\cite{vasarhelyi2020computational,spoon2023gender}.

Teamwork can also involve an unequal allocation of credit among members. Empirical evidence from the academic field of economics indicates that 
while men and women who solo-author most of their papers exhibit similar career outcomes, 
co-authored papers are more strongly associated with successful career outcomes for men than for women~\cite{sarsons2017recognition}. 
Experimental evidence suggests that this effect might be due to stereotyped credit attribution~\cite{sarsons2021gender}. Further, articles written by female authors are more likely to
be omitted from a reference list~\cite{koffi2021innovative}. Taken together, the above findings indicate that women must often navigate a more complicated landscape to achieve success.

Inequalities in success and their structural underpinning in networks and teams suggest that the gender gap in success is unlikely to close in the near future. In fact, the COVID-19 pandemic reversed many of the earlier advances that made the playing field more level. After the pandemic began, millions of women dropped out of the workforce globally~\cite{deloitte2021women}, with the greatest numbers among women with young dependents~\cite{myers2020unequal}. More research and interventions are needed to bring greater awareness to this challenge. Identity-based disparities in success outside of binary cisgender norms and in terms of race, ethnicity and disability are likely to be even more pronounced, calling for future work in this critical area.

\section{Outlook}

Taken together, this review highlights how key dimensions of success can be understood as collective phenomena emerging from interactions of interconnected individuals. By viewing success through a complex systems lens, we can better understand its dynamics and underlying mechanisms. 
While our focus has been on aspects of success linked to collective dynamics,  
as illustrated in Fig.~\ref{fig:framework2}, other areas---such as individual psychology and performance, consumer psychology, and qualitative research in management science---remain outside the scope of this review. Future research could integrate this vast body of literature with the effects reviewed here.

This review details remarkable regularities and predictive insights that, in some cases, can be generalized across domains.
It also highlights studies that have challenged the generality of long-standing paradigms by revealing important contextual factors that delimit their validity boundaries.
For example, the canonical view that social hubs facilitate the large-scale adoption of new products and behaviors--in the same way that they act as superspreaders of infectious diseases--does not hold for complex contagions that require social reinforcement~\cite{watts2007influentials,guilbeault2021topological}.
The informational advantages from both weak and strong social ties are more nuanced than they first appeared and depend on the context in which the individual is embedded~\cite{aral2011diversity,aral2022exactly,rajkumar2022causal}. 
Similarly, the common belief that early successes are more likely than early failures to lead to later success does not always hold, especially when individuals learn from failure and improve their efforts~\cite{wang2019early,yin2019quantifying}. The contextual factors in these examples challenge the notion that there might be some universal recipes for success. This realization does not imply that no useful theories can be built, but -- akin to other social phenomena~\cite{almaatouq2022beyond} -- it calls for more research on understanding the scope and boundary conditions for various theories of success.

Among the contextual factors that affect success, cultural and geographical differences may be important to explore in future research.
Marketing scholars have long understood that the spread of new products and behaviors varies significantly among countries -- even for the same products -- which has been explained by both cultural factors and economic factors~\cite{peres2010innovation}. 
In a different stream, organizational behavior research find that cultural differences affect individuals' definitions of career success~\cite{benson2020cultural}, as well as the correlates of career success~\cite{smale2019proactive}.
The managerial literature on multicultural teams reports that cultural diversity improves team creativity and problem-solving ability even though it may also cause misunderstanding, undermine trust, and increase conflict~\cite{stahl2021unraveling}. 
Future research could integrate cultural considerations into the insights reviewed here. 

Other essential directions for future research include work on inequality, interventions, and algorithms. While strong regularities such as success-breeds-success effects and systematic biases may facilitate accurate predictions, these phenomena may also lead to greater inequality and undermine the relationship between success and merit. We need additional research on inequality to learn more about the link between success and demographic differences, including gender, race, ethnicity, nationality, age, and class, in various settings where these factors could play a role.
Future research could further improve our understanding of the economic and ethical importance of tackling issues of social inequality, which will help frame discussions aimed at bringing pressing, inequality-related problems to policymakers. For instance, we need to systematically uncover biases against ideas that primarily benefit historically disadvantaged groups~\cite{nielsen2018making}.
A great deal of work also remains on developing approaches to mitigate social inequality in traditional as well as AI-mediated settings. 

Interventions present another promising avenue for research on success, raising key questions: Can we use our research insights to help ideas, people and organizations succeed? Can we apply our insights about success through large-scale field experiments to facilitate better outcomes for individuals and teams and for the ideas and products they create?
Although such experiments pose logistical challenges and carry potentially significant ethical concerns due to unintended consequences, the benefits could be substantial. 
Indeed, these experiments could not only improve the causative interpretations of many of the insights about success reviewed in this article, but also facilitate the design of more diverse, inclusive, and sustainable social systems in which individuals have the same opportunities to succeed regardless of their gender and socioeconomic status; sustainable behaviors constitute established social norms; and ideas are judged based on their merit and not the prestige of their creators and their institutions. Research progress in this direction could especially help traditionally underrepresented groups and those in developing regions of the world.

More research is also needed on algorithms and their long-term implications. Increasingly, success is shaped by algorithms across many domains, and these algorithms are growing more and more sophisticated. Algorithms can introduce new biases that are difficult to detect. We need to better understand how algorithms influence success to ensure fairness and inclusion~\cite{mehrabi2021survey}. Algorithm sophistication is not the only problem. In performance evaluation and ranking, even simple metrics, when applied acritically, may cause unintended strategic behavioral responses by individuals.
In academia, for example, the use of citation-based metrics in research evaluation has been associated with questionable practices~\cite{hall2019towards,west2021misinformation}, such as self-citations by individuals~\cite{hall2019towards} and journals~\cite{siler2022games}, citation cartels~\cite{kojaku2021detecting}, and predatory publishing~\cite{west2021misinformation}.
Behavioral responses to the deployment of metrics for decision-making can not only invalidate the metrics as good proxies for individuals', teams' and organizations' performance, as predicated by Goodhart's law~\cite{strathern1997improving}, but they can also stifle originality and collective progress~\cite{azoulay2011incentives}.
Understanding ways to prevent the unintended consequences of quantitative evaluation algorithms represents a complex research challenge, which requires further cross-disciplinary efforts.

Finally, this review has focused primarily on well-established dimensions of success, driven by the overall direction of the literature.
It has not addressed the normative question of how success ought to be defined for individuals and society more broadly. In academia, for example, publication in leading journals is viewed by some as a measure of success. Yet, evidence from academic papers in political science finds that leading journals publish women-dominated topics at lower rates than they do for man-dominated ones~\cite{key2019you}. Further, data on US PhD recipients between 1980 and 2010 indicates that women studying topics associated with women have fewer career opportunities~\cite{kim2022gendered}.
While reducing gender biases represents an important step toward a more equitable society, aligning success with merit remains challenging, partly because the quantification of merit is often elusive. Definitions of merit based on an individual’s talents and skills may neglect the extent to which the individual uses her talents for actions that have positive consequences for the system~\cite{sen2000merit}. On the other hand, definitions based on the outcomes of an individual’s actions depend on evaluations of success and failure, which do not always reflect societal values and can potentially perpetuate inequalities~\cite{sen2000merit}.
Some studies have already started identifying the conditions under which we see a decoupling of merit and success--e.g., through agent-based models and experiments~\cite{salganik2006experimental,van2019self}-- suggesting ways to prevent those conditions. These studies, however, rely on simplified scenarios in which a single dimension of merit is clearly identifiable, which is rarely the case in real social systems. More research is needed to understand how to align success, merit, and societal values, ultimately creating more equitable societies.

In the long run, we envision that these efforts will shift the focus of success research from understanding and predicting success to designing more effective, sustainable, diverse, and inclusive social systems. Achieving this vision will require researchers from diverse fields to collaborate and integrate a wide range of ideas and methodologies. Such collective work could pave the way toward social systems where success better reflects quality, talent, and societal values--and where everyone has equal opportunities to flourish, regardless of their backgrounds.

\clearpage

\backmatter

\bmhead{Acknowledgements}

MSM acknowledges financial support from the URPP Social Networks of the University of Zurich and the Swiss National Science Foundation (Grant No. 100013-207888). 
This version of the article has been accepted for publication, after peer review but is not the Version of
Record and does not reflect post-acceptance improvements, or any corrections. The Version of Record of this article is published in \textit{Nature Communications}, and
is available online at \url{https://doi.org/10.1038/s41467-024-54612-4}.

\bmhead{Author Contributions Statement}

MSM and DW conceived this review article and designed its structure; MSM and DW wrote the first draft of the introduction; MSM wrote the first draft of the section on the collective nature of success and outlook; MSM, FB, EAH, GL, FM wrote the first draft of the remaining sections; MSM, EAH and DW revised the whole text.



\bmhead{Competing Interests Statement}

The authors declare no competing interests.

\clearpage



\end{document}